\documentclass{ceab}    

\usepackage{epsfig}     
\usepackage{graphicx}   
\usepackage{ceabbib}    
\usepackage[T1]{fontenc}

\begin{document}

\def\tit{Multiplicity of high-mass stars}
\def\aut{R. Chini et al.}
\def\str{??--??}
\def\gore{R. Chini et al.}

\title{THE MULTIPLICITY OF HIGH-MASS STARS}

\author{R. Chini$^{1,2}$, A. Barr$^1$, L.~S. Buda$^1$, T. Dembsky$^1$,  H. Drass$^1$,A. Nasseri$^1$, \\ V.~H. Hoffmeister$^1$, and K. Fuhrmann$^{1,3}$
\vspace{2mm}\\
\it $^1$Astronomisches Institut, Ruhr-Universit\"at Bochum,\\ 
\it Universit\"atsstra\ss{}e 150, D-44780 Bochum, Germany\\
\it $^2$Instituto de Astronom\'{i}a, Universidad Cat\'{o}lica del Norte,\\ 
\it Antofagasta, Chile\\
\it $^{3}$Isaac Newton Group of Telescopes, Apartado 321,\\ 
\it E-38700 Santa Cruz de La Palma, Spain\\
}

\maketitle

\begin{abstract}
We report about an ongoing photometric and spectroscopic monitoring survey of about 250 O- and 540 B-type stars in the southern Milky Way with the aim to determine the fraction of close binary systems as a function of mass and to determine the physical parameters of the individual components in the multiple systems. Preliminary results suggest that the multiplicity rate drops from $\sim 80\%$ for the highest masses to 20\% for stars of 3 $M_\odot$. Our analysis indicates that the binary systems often contain close pairs with components of similar mass. This coincidence cannot originate from a random tidal capture in a dense cluster but is likely due to a particular formation process for high-mass stars. The large percentage of multiple systems requires a new photometric calibration for the absolute magnitudes of O-type stars.
\end{abstract}


\section{Introduction}

The most massive stars, historically classified as O-type stars, are rare objects with masses above 16\,M$_\odot$, typically found at large distances from the sun; early B stars ($M > 8\,$M$_\odot$) also contribute to the group of high-mass objects. The formation of these O- and B-type stars is still under debate and can be explained by various models (Zinnecker \& Yorke 2007). Both observations of massive circumstellar disks (Shepherd, Claussen \& Kurtz 2001; Chini et al. 2004; Cesaroni et al. 2005; Patel et al. 2005; Kraus et al. 2010) and theoretical calculations (Yorke \& Sonnhalter 2002; Krumholz et al. 2009; Kuiper et al. 2010) seem to favour the accretion scenario. However, the high multiplicity among high-mass stars (Preibisch et al. 1999; Mason et al. 2009) might alternatively support a merging process of intermediate-mass stars (Bonnell, Bate \& Zinnecker 1998).

Massive binary stars are believed to be the progenitors of a variety of astrophysical phenomena, e.g. short gamma-ray bursts (Eichler et al. 1989; Paczynski 1991; Narayan, Paczynski \& Piran 1992), X-ray binaries (Moffat 2008), millisecond pulsar systems and double neutron stars (van den Heuvel 2007). Even more relevant, the multiplicity of stellar systems is a crucial constraint for the various star formation scenarios, particularly if the multiplicity fraction were a function of stellar mass. While many massive stars are found to be part of binary or multiple systems, comprehensive statistics on close binaries is still missing. The smallest separations are expected to be around 0.2 AU, resulting in orbital periods of a few days only. Below this minimum distance the binary system will merge to form a single object.

The vast parameter space in possible orbital periods and mass ratios requires different, partly overlapping complementary methods that have their own limitations and their observational biases (Sana \& Evans 2011): High-resolution imaging like adaptive optics or interferometric techniques serve for systems with wider separations and mass ratios $q = M_2 / M_1$ between 0.01 and 1 while high-resolution spectroscopy is biased toward finding close companions and those that are a significant fraction of the primary's mass ($q > 0.1$). We note that the inclination of the orbit with respect to the observer and the eccentricity are other crucial limitations for the spectroscopic detection of close multiple systems.

An adaptive optics survey of about a third of the known galactic O stars revealed visual companions in 27\% of the cases (Sana \& Evans 2011). Speckle interferometry for most of the galactic O stars showed companions for 11\% of the sample (Mason et al. 2009). Moreover, the same study claimed that 51\% of the O-type objects are spectroscopic binaries (SBs), based on an extensive review of the literature. This is in accord with a recent spectroscopic survey finding that among $\sim 240$ southern galactic O and WN stars more than one hundred stars show radial velocity ($RV$) variations larger than 10\,km/s (Barb\'{a} et al. 2010). Recently, a high-resolution imaging campaign of 138 fields containing at least one high-mass star yielded a multiplicity fraction close to 50\% (Ma\'{i}z-Apell\'{a}niz 2010). In summary, the spectroscopic binary frequency of high-mass stars so far observed and reported in the literature is moderately high, while the visual binary fraction is low. Little or no discussion as to why the multiplicity for O-type stars is high has been offered up to now nor have conclusions been drawn so far about which of the competing high-mass star formation models best explains the hitherto known trends in the stars' multiplicity.

There have been several surveys of B star duplicity in the past. In a sample of 109 B2 - B5 stars there were 32 (29\%) spectroscopic and 49 (45\%) visual binaries yielding a total binary frequency of 74\% (Abt, Gomez \& Levi 1990). A spectroscopic survey of 83 late B-type stars revealed that 24\% of the stars had companions with mass ratios greater than 0.1 and orbital periods less than 100 days (Wolff 1978). A speckle interferometry survey of the Bright Star Catalogue resolved 34 of 245 B stars into binaries corresponding to a multiplicity fraction of ~14\% (McAlister et al. 1987, 1993).

Another speckle interferometry survey of 48 Be stars revealed a similar binary fraction of $10\% \pm 4\%$ (Mason et al. 1997). A further comparison study, based on adaptive optics IR imaging and probing separations from 20 to 1000 AU for 40 B and 39 Be stars derived the same qualitative result, i.e. that the multiplicity of B and Be stars are identical. This time, however, the binary fractions were $29\% \pm 8\%$ for the B stars and $30\% \pm 8\%$ for the Be stars (Oudmaijer \& Parr 2010). Finally an adaptive optics photometry and astrometry survey of 70 B stars revealed 16 resolved companions (23\%) (Roberts, Turner \& Ten Brummelaar 2007). In summary, the overall multiplicity fraction of high-mass stars seems to decrease with stellar mass.

In the present paper we investigate the multiplicity fraction in the stellar mass range of about $3 - 80\,M_\odot$ and for mass ratios $q > 0.2$ and search for new eclipsing O-type binaries.

\section{The Survey}

\begin{figure}
\begin{center}
	\includegraphics*[width=100mm]{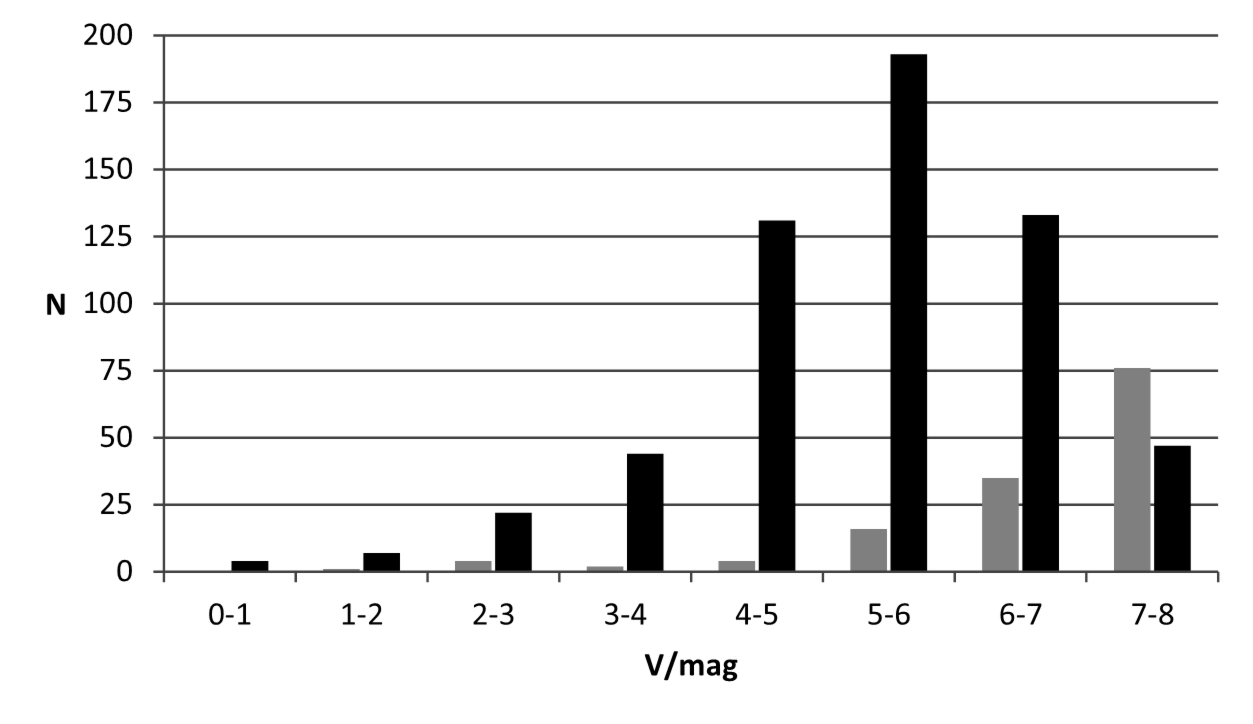}
	\caption{
    		Distribution of $V$ magnitudes for the samples of O-type stars (grey) and
            B-type stars (black) columns. The B stars are on average brighter which results in better $S/N$ spectra and thus increases the chance for detecting fainter companions.
			}
\label{histo_mag}
\end{center}
\end{figure}

 Our spectroscopic survey comprises 138 O- and 581 B-type stars with $V \le 8$\,mag. The O-stars were taken from the Galactic O-Star Catalogue V.2.0 (Sota et al. 2008) (GOSC). The B stars were selected from the HIPPARCOS archive: 50\% of the B stars form a volume-limited sample with $d < 125$\,pc, the remaining 50\% have been chosen to provide roughly an equal amount of stars for each subclass from B0 to B9. The distribution of visual magnitudes is displayed in Fig.~\ref{histo_mag} showing that the B-star sample contains on average brighter stars than the O-star sample.

\subsection{spectroscopy}

Using the high-resolution spectrograph BESO (Fuhrmann et al. 2011) at the Hexapod Telescope at the Universit\"atssternwarte Bochum near Cerro Armazones in Chile we obtained 4577 multi-epoch optical spectra. The observing period started in January 2009 and is still going on. The spectra comprise a wavelength range from 3620 to 8530\,\AA\, and provide a mean spectral resolution of $R = 50,000$. The entrance aperture of the star fibre is $3.4"$. The integration time per spectrum was adapted to the published visual brightness of each star. It was our primary goal to monitor a large number of stars rather than to obtain a very high $S/N$ for individual stars. As a consequence of this strategy potential companions fainter than $\Delta V \sim 2$\,mag are barely visible in our spectra, thus decreasing the chance for detecting SB2s. Converting this brightness difference of 2\,mag into a mass difference we are sensitive to mass ratios $q > (0.18 -  0.40)$ for O5 - O9 stars and $q > (0.43 - 0.55)$ for B stars. In other words, the detectable companions of an O5 star range from O5 to about B2, those of a B9 star from B9 to about A7.

Additional spectra were collected during various observing runs with FEROS at the ESO 1.5\,m and the MPG/ESO 2.2\,m telescopes, both located on La Silla, Chile. Because BESO is a clone of FEROS the instrumental parameters are identical to those described above. These spectra cover a time span between 2006 and 2008.

\begin{figure}[h]
\begin{center}
	\includegraphics*[width=100mm]{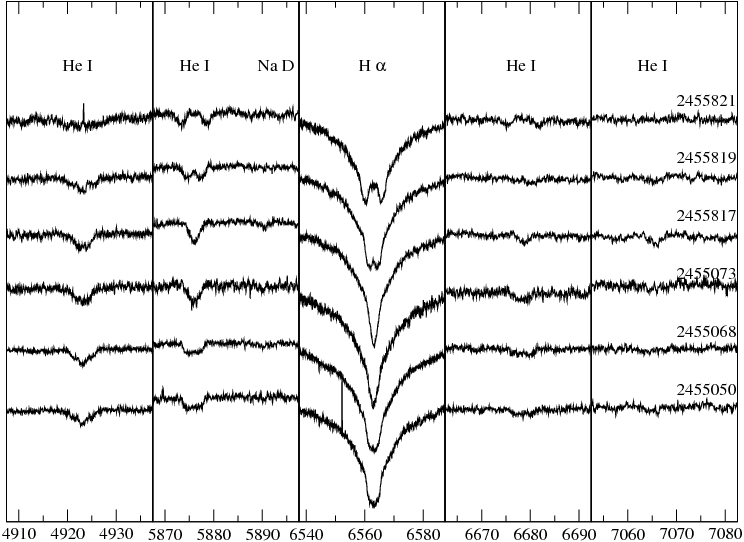}
	\caption{
			Multi-epoch BESO spectra obtained for the so far unknown double-lined B star binary HD\,22203 (19\,Eri) between JD\,$2455050 - 2455821$. The selected spectral regions show H$\alpha$ and several He\,I lines. The change from a single- to a double-lined structure is clearly visible.
			}
\label{HD22203}
\end{center}
\end{figure}

Finally we extracted 824 O-star spectra from the ESO archive covering an observing period from 1999 to June 2011. These spectra were also obtained with FEROS and processed in the same way as the BESO spectra.

Until today there are 1849 O- and 2728 B-star spectra in our archive. For each O-star we collected about ten spectra on the average with at least two spectra for those objects which were known to be multiple systems before and up to of 39 spectra in such cases where multiplicity was not immediately obvious. The spectra are separated by days, weeks and months. So far we have observed 550 stars from our B-type sample with an average number of five spectra per star. We stopped temporarily collecting data for those B-stars where the multiplicity became obvious during the first two spectra.

All data were reduced with a pipeline based on the MIDAS package developed for the ESO FEROS spectrograph. A quantitative analysis was done with standard IRAF line-fit routines allowing the detection of line shifts in single-line spectroscopic binaries (SB1) or line deformation and/or separations in double-line spectroscopic binaries (SB2). For the O- and early B-stars we used exclusively He lines for identification; He\,I ($\lambda 5875$\,\AA) was particularly useful due to the nearby interstellar Na\,I doublet ($\lambda\lambda  5890, 5896$\,\AA) allowing for a sensitive verification of any relative line shift. For later B-types we had to rely primarily on hydrogen lines. A typical set of multi-epoch spectra is shown in Fig.~\ref{HD22203}.

\subsection{Photometry}

\begin{figure}
\begin{center}
	\includegraphics*[width=84mm]{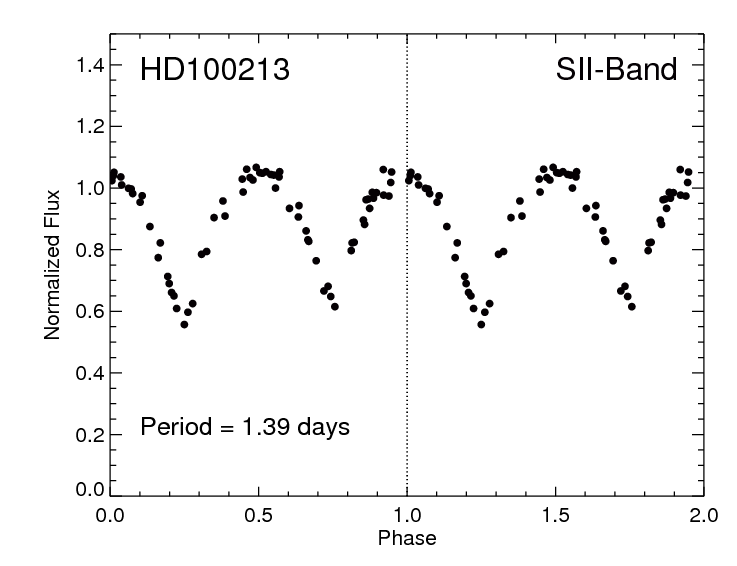}
	\caption{
			Light curve for HD\,100213, a previously known eclipsing binary with a
            period of 1.39 days (Linder et al. 2007).
			}
\label{EB_HD100213}
\end{center}
\end{figure}

Simultaneously with our spectroscopic survey we are performing a photometric monitoring at the robotic VYSOS\,6 telescope (Haas et al. 2012) also hosted at the Universit\"atssternwarte Bochum. Due to the high brightness of the stars we typically use $I$- or narrow-band filters to avoid saturation. In a first approach we have concentrated on the O-star sample with the goal to obtain $30 - 50$ photometric values in consecutive nights -- a strategy which is obviously biased against periods shorter that one month. In a second photometry run we will follow those stars that show slow variations suggesting longer periods. To give an example for the current quality of our data we show the light curve for HD\,100213 in Fig.~\ref{EB_HD100213}. We obtain a period of $P = 1.39$ days which perfectly agrees with the results by Linder et al. (2007).

\section{Results}

\subsection{Spectroscopy}

\begin{figure}[h]
\begin{center}
	\includegraphics*[width=100mm]{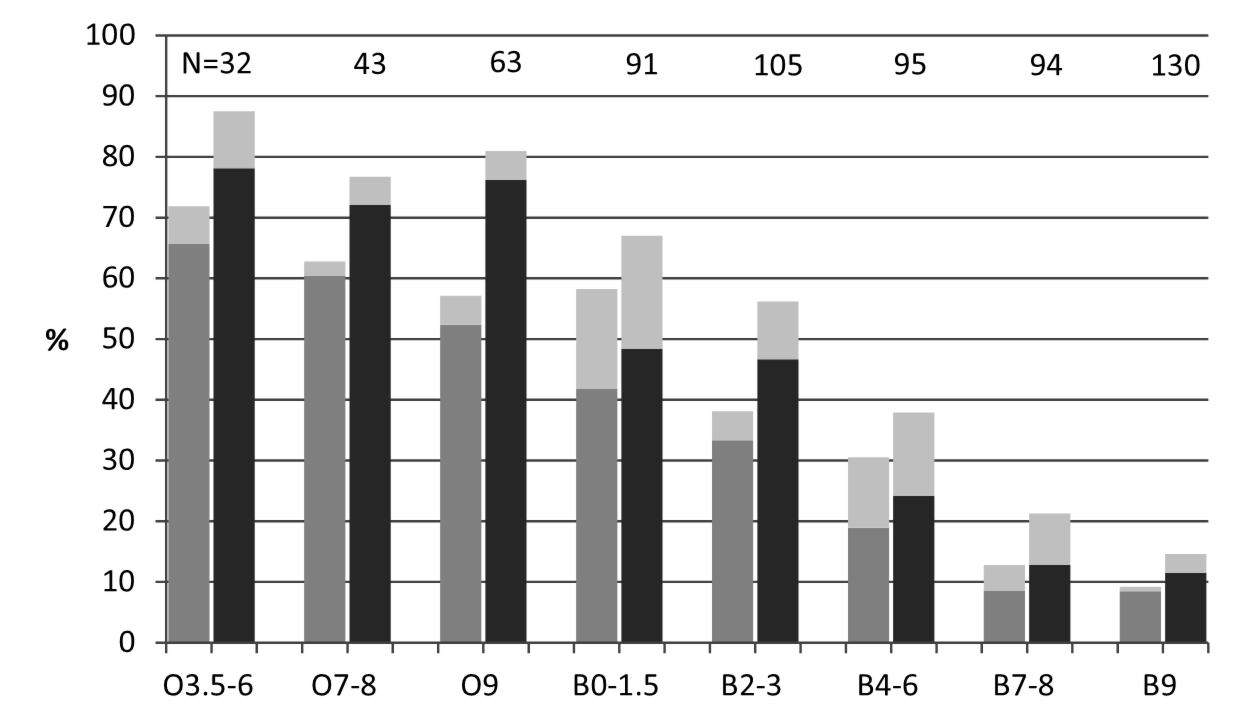}
	\caption{
			Multiplicity vs. spectral type of O- and B-type stars. The black columns
            denote the total percentage of spectroscopic binaries, the grey columns display the percentage of double-lined spectroscopic binaries. The number \emph{N} on top of the columns denote the number of objects in each spectral bin. The shaded areas on top of each column show further potential candidates for $RV$ variability.
			}
\label{histo_mult}
\end{center}
\end{figure}

We detect \emph{RV} variations for about 75\% of all O stars brighter than $V = 8$\,mag while additional 6\% are potential candidates for variable \emph{RV}. For B0 stars the variability fraction is 48\% (67\%)  decreasing to 12\% (15\%) for B9 stars (Fig.~\ref{histo_mult}). In general, our observed multiplicities are lower limits due to the unknown orbit inclination and the constraints for the detectable minimum mass ratio.

The overall percentage of SBs for O stars is higher than found before in similar investigations and is due to the numerous spectra obtained for a single star, which reveal mainly variations within days and weeks that were not obvious in previous studies. As mentioned above Mason et al. (2009) report a general spectroscopic multiplicity fraction of 51\% for the population of Galactic O stars. Individual nearby clusters were found to have binary fractions between 0\% and 63\% with an average value of 44\% (see e.g. Sana \& Evans 2011 for an overview).

Inspecting those 60 O stars that did not show any $RV$ variations throughout the last six years, we find that additional 13 out of 29 stars observed through adaptive optics measurements (Mason et al. 1997) or speckle interferometry (McAlister et al. 1993) possess visual companions. These complementary data increase the total percentage of multiple systems for stars brighter than $V = 8$\,mag to 91\% leading us to suggest that basically all O-stars are members of multiple systems. This finding also goes beyond the results summarized by Sana \& Evans (2011) who obtain a total minimum multiplicity fraction close to 70\%.

Among the SBs in the present study 65\% (72\%) of the spectra for O stars contain more or less separated multiple lines (SB2) reflecting that the majority of systems contains pairs of similar mass. This is in agreement with results by Kobulnicky \& Fryer (2007) who found that massive stars preferentially have massive companions. From our current time coverage, however, it is not yet possible to derive orbital periods and constrain semi-amplitudes of radial velocities and hence accurate binary component mass ratios.

Although the general multiplicity fraction seems to decrease with spectral type the relative high fraction of SB2 compared to SB1 remains valid until the latest B-types. Due to the limited spectroscopic binary studies for B stars a comparison between our work and previous investigations is difficult. Kouwenhoven et al. (2005) studied the binary population in the Scorpius OB2 association. Summarizing the results from all techniques they find multiplicity fractions of $\sim 80\%$ and $\sim 50\%$ for B0 - B3 and B4 - B9, respectively. While the numbers for the early B-types are compatible with our study the multiplicity rate for the late B-types is slightly higher in Sco OB2; the latter difference might be due to the different observing techniques, because without the adaptive optics data of Kouwenhoven et al. (2005) the multiplicity rate would drop by about 10\% for the late B-types. Kouwenhoven et al. (2007) repeated the analysis of the primordial binary population in Sco OB2: one set of spectroscopic data in this study comes from Levato et al. (1987) who found that the binary fraction is at least 30\% for all early-type stars but might be as high as 74\% if all reported \emph{RV} variations were due to binaries. Another spectroscopic study (Brown \& Verschueren 1997) included in the analysis by Kouwenhoven et al. (2007) yields a similar range between 28\% and 76\%. However, the dependence on stellar mass has not been addressed in these studies.

Miroshnichenko (2010) investigated the multiplicity for bright galactic Be stars and found that their binary fraction should be at least 50\% independent of the spectral type. In contrast, the AO study of Oudmaijer \& Parr (2010) yielded a binary fraction of only $30\% \pm 8\%$ for 39 Be stars. The same authors claim that the binarity of normal B stars is $29\% \pm 8\%$ and thus identical to those of Be stars. Wheelwright, Oudmaijer \& Goodwin (2010) investigated the binarity of 25 Herbig Be stars with spectro-astrometry and derive a high binary fraction of 74\%. Obviously, the multiplicity of B stars is still a subject that needs further investigation; the existing results -- including ours -- likely suffer from various biases. We actually might expect to observe more binary stars amongst late type B stars compared to O stars, however, then their mass ratios must be larger than what is covered by our study.

\subsection{Photometry}

\begin{figure}
\begin{center}
	\includegraphics*[width=84mm]{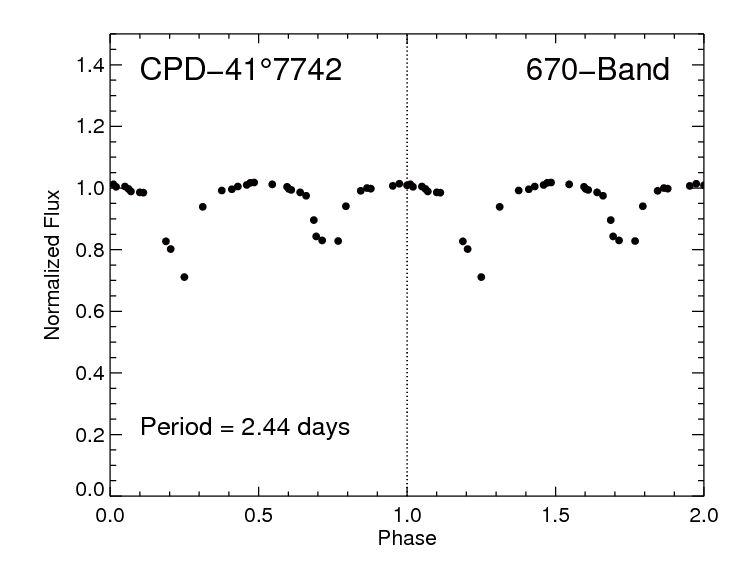}
	\caption{
			Light curve of CPD$-41\,7742$, an early-type spectroscopic binary in the young open cluster NGC6231,
            obtained through a 670\,nm narrow band-filter. We derive a photometric period of $P =2.44$ days
            in agreement with the spectroscopic result by Sana et al. (2003).}
\label{EB_CPD417742}
\end{center}
\end{figure}

Currently our analysis comprises 233 (out of 249) O-type stars from the GOS catalogue. We found variability for 56 stars corresponding to about 24\% of the sample. Lef\`{e}vre et al. (2009) argue that 17\% of all high-masss stars with magnitudes $V < 8$ can be considered as eclipsing binaries.  We could verify the orbital periods of four previously known eclipsing binaries: CPD$-59\,2603$ (Rauw et al. 2001), HD\,093206 (Walker \& Merino 1972), HD\,100213 (Terrel et al. 2003), and HD\,152219 (Sana et al. 2006). Fig.~\ref{EB_CPD417742} shows the light curve for CPD$-41\,7742$, an early-type spectroscopic binary in the young open cluster NGC\,6231; we obtain a period of $P = 2.44$ days in agreement with the spectroscopic result by Sana et al. (2003).

Furthermore, there are three stars whose multiplicity and orbit period were already known due to spectroscopic studies but for which a photometric light curve is still missing: CPD$-59\,2635$  (Colombo et al. 2001), HD\,152590 (Gieseking 1982), and HD\,101205 (Balona 1992). Eventually there are ten stars which are classified as eclipsing or spectroscopic binaries in the literature and that show variability in our photometric data. However, the number of observations is still not sufficient to establish an orbital period: CPD$-59\,2641$ (Luna et al. 2003), Tau\,CMa (van Leeuwen \& Van Genderen 1997), $\delta$\,Cir (Penny et al. 2001), HD\,153919 (Rubin et al. 1996), HD\,152218 (Struve 1944), CPD$-41\,7733$ (Sana et al. 2007), HD\,150136 (Niemela \& Gamen 2005), HD\,165052 (Arias et al. 2002), and HD\,159176 (Lloyd Evans 1979).

The fact that most O-type stars occur as binaries suggests that the existing photometry is contaminated and thus their absolute magnitude calibration has been overestimated; in the worst case the tabulated O star magnitudes are wrong by 0.75\,mag at each wavelength. We aim at a new photometric calibration for O-type stars on the basis of our spectroscopic and photometric surveys.

\section{Discussion}

\begin{figure}
\begin{center}
	\includegraphics*[width=100mm]{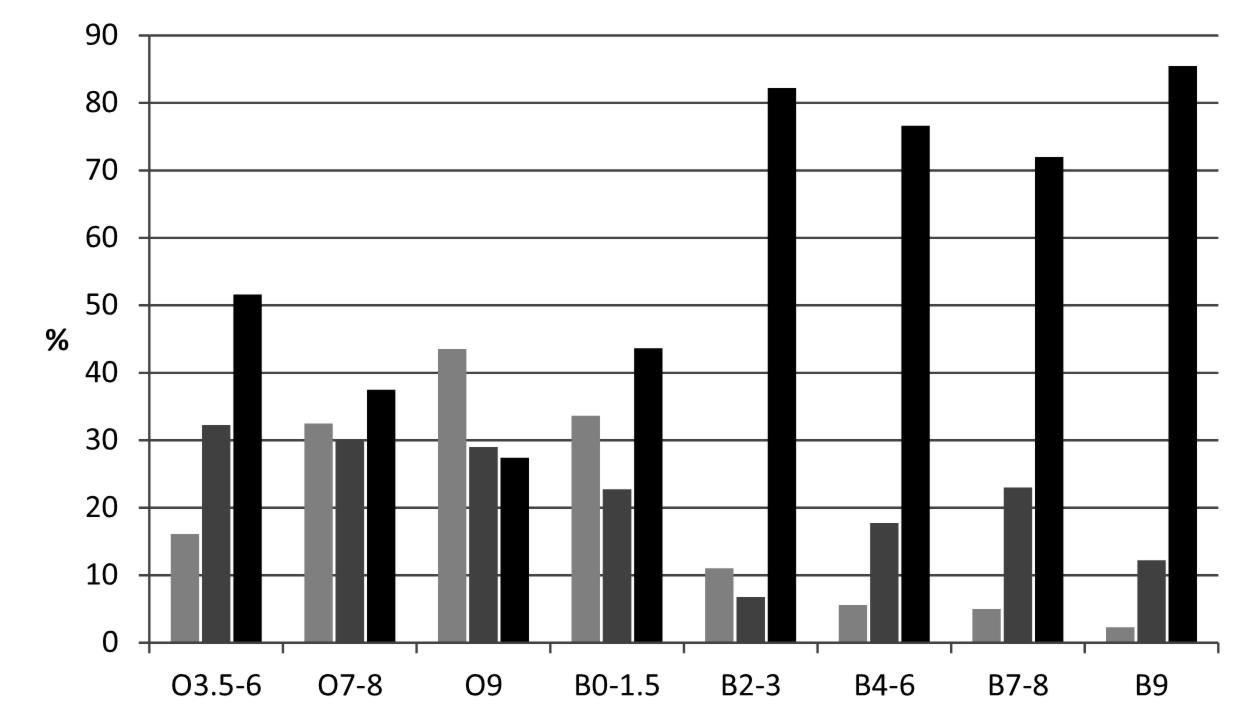}
	\caption{
			Luminosity classes for each spectral bin.
			}
\label{histo_evo}
\end{center}
\end{figure}

Fig.~\ref{histo_mult} shows a clear trend for the multiplicity to decrease with mass. Future observations will show whether part of this trend is due to our observing strategy and due to the fact that the number of useful spectral lines decrease from early O to late B. One may also speculate whether the decrease of the binarity fraction towards lower masses is due to an \emph{evolutionary} effect: an O-type star -- whether on the main sequence or not -- is much younger than a late B-type star on the main sequence and is likely more evolved due to its rapid hydrogen burning. If evolution "destroys" binarity then evolved stars (i.e. luminosity classes III and I) should have lower binarity fractions than stars on the main sequence.  Fig.~\ref{histo_evo} shows the distribution of luminosity classes I, III, and V for the different spectral type bins in our sample. Obviously those spectral types with the highest binarity fraction (O3 -- B1) have the highest fraction of evolved stars while late B-Type stars with the lowest binarity fraction comprise only few evolved stars. This excludes the possibility that evolution influences binarity. On the other hand, one can exclude a pure \emph{age} effect as SB2 binaries are all very tight and hard to break up. They could dynamically interact in dense cluster cores and get ejected, but breaking up such "hard" SB2s is not very likely.

The multiplicity of high-mass stars seems to depend on their environment. It was found that the binary frequency among O stars in clusters and associations is much higher than among field stars (which have no apparent nearby cluster) and runaway stars (O stars with peculiar $RV$s in excess of 40\,km/s or remote from the Galactic plane). The spectroscopic binary fraction in clusters and associations, in the field, and among runaways obtained in two different studies (Gies 1987; Mason et al. 1998) were 55 (61)\%, 45 (50)\% and 19 (26)\%, respectively. For the classification of the environment we have used the designations in the GOSC (Sota et al. 2008) refined by a recent investigation which investigated the origin of field O-stars (Schilbach \& R\"oser 2008). The fractions of SBs within the individual groups are: clusters 11\% ($N = 15$), associations 59\% ($N = 82$), field 6\% ($N = 8$), and runaways 21\% ($N = 29$); we have only counted the secure classifications (black colums in Fig.~\ref{histo_mult}. Our results for clusters is based on rather low numbers. Taking the average of clusters and associations our results are compatible with previous studies (Gies 1987; Mason et al. 1998). The same holds for our binary fraction among the runaway stars; Gies (1987) obtained $19 \pm 5\%$ while Mason et al. (1998) found $26 \pm 5\%$, respectively.

While the direct observation of disks around the earliest O stars will remain a fundamental challenge to test model computations for individual high-mass systems, binary star statistics offers other constraints on hydrodynamical simulations of star forming clusters. Early calculations of the fragmentation of an isothermally collapsing cloud have already shown that binary systems and hierarchical multiple systems are frequently obtained (Larson 1978). Although the variety of processes and their sequences are manyfold (Bate, Bonnell \& Bromm 2002), there is consensus that the accretion processes favor the formation of binaries with mass ratios $q \sim 1$ for those systems with separations below 10\,AU, (e.g. Clarke (2007) and references therein).

Nevertheless, the simulations almost never produce binaries with $q < 0.5$. This appears to be a general problem of turbulent fragmentation calculations which seems to origin from gas accretion onto a proto-binary (Goodwin, Whitworth \& Ward-Thompson 2004). Independent of how the binary system was formed, the infalling material has a high specific angular momentum compared to the binary and thus will preferentially accrete onto the secondary (Bate et al. 2002). Bate (2000) even predicts that binaries will evolve rapidly towards $q \sim 1$ regardless of their initial mass ratio. For the special case of massive stars recent theoretical calculations claim that close high-mass stellar twins can in principle be formed via fragmentation of a disk around a massive protostar and subsequent mass transfer in such close, rapidly accreting oversized proto-binaries (Krumholz \& Thompson 2007). This scenario provides a natural explanation for the numerous high-mass spectroscopic binaries with mass ratios close to unity. Most likely, mass transfer will continue during the entire life of such a close system leading to continuously changing properties of the individual components and thus to a different evolution compared to a single star.

The high binary frequency for clusters found in the present study exclude random pairing from a classical IMF as a process to describe the similar-mass companions in massive binaries. Our results make binary-binary interactions inside clusters very probable and thus can explain the high binary fraction among runaway stars. Because ejection requires a cluster origin of the binaries \citep{K01} it supports the model of competitive accretion within the cluster environment. Finally, the small number of field O stars - only four "certified" stars remain currently in this category - suggests that probably all O stars are born in associations or clusters.

Radiation-hydrodynamic simulations show that, during the collapse of a massive prestellar core, gravitational and Rayleigh-Taylor instabilities channel gas onto the star system through non-axisymmetric disks. Gravitational instabilities lead to a fragmentation of the disk around the primary star and form a massive companion. Radiation pressure does not limit stellar masses, but the instabilities that allow accretion to continue lead to small multiple systems (e.g. Krumholz et al. 2009).

The current study was a pure discovery project that aimed exclusively at the multiplicity statistics in the mass range $3 < M \rm{[}M_\odot\rm{]} \simeq 80$. In a next step we will study the orbital properties and the spectral types of the individual components to obtain a statistically relevant archive for high-mass binary systems.

\section*{Acknowledgements}
We wish to thank the numerous colleagues from the Ruhr-Universit\"at Bochum for their help with the observations and we acknowledge the continuous support by the Universidad Cat\'{o}lica del Norte in Antofagasta. This publication is supported as a project of the  Nordrhein-Westf\"alische Akademie der Wissenschaften und der K\"unste in the framework of the academy program by the Federal Republic of Germany and the state Nordrhein-Westfalen.

\bibliographystyle{ceab}
\bibliography{sample}

\section*{References}
\begin{itemize}
\small
\itemsep -2pt
\itemindent -20pt

\item[] Albacete Colombo, J.~F., Morrell, N.~I., Niemela, V.~S., Corcoran, M.~F. 2001, \mnras, 326, 78
\item[] Arias, J.~I., Morrell, N.~I., Barb{\'a}, R.~H., et al., 2002, \mnras, 333, 202
\item[] Balona, L.~A., 1992, \mnras, 254, 404
\item[] Chini, R., Hoffmeister, V.~H., Nasseri, A., Stahl, O., and Zinnecker, H.: 2012, \mnras, 424, 1925.
\item[] Abt H.A., Gomez A.E., Levi S.G., 1990, \apjs 74, 551
\item[] Barb\'{a} R.H., Gamen R., Arias J.I., Morrell N., Ma\'{i}z-Apell\'{a}niz J., Alfaro E., Walborn N., Sota A., 2010, RMxAA 38, 30
\item[] Bate M.R., 2000, \mnras 314, 33
\item[] Bate M.R., Bonnell I.A., 1997, \mnras 285, 33
\item[] Bate M.R., Bonnell I.A., Bromm V., 2002, \mnras 336, 705
\item[] Bonnell I., Bate M.R., Zinnecker H., 1998, \mnras 298, 93
\item[] Brown A.G.A., Verschueren W., 1997, \aap 319, 811
\item[] Cesaroni R., Neri R., Olmi L., Testi L., Walmsley C.M., Hofner P., 2005, \aap 434, 1039
\item[] Chini R., Hoffmeister V.H., Kimeswenger S., Nielbock M., N\"urnberger D., Schmidtobreick L., Sterzik M., 2004, Nat. 429, 155
\item[] Clarke C.J., 2007, IAUS 240, 337
\item[] Eichler D., Livio M., Piran T., Schramm D.N., 1989, Nat. 340, 126
\item[] Fuhrmann K., Chini R., Hoffmeister V.H., R.Lemke, Murphy M., Seifert W., Stahl O., 2011, \mnras
\item[] Gies D.R., 1987, \apjs 64, 545
\item[] Gieseking, F.\ 1982, \aaps, 49,
\item[] Goodwin S.P., Whitworth A.P., Ward-Thompson D., 2004, \aap 423, 169
\item[] Kobulnicky H.A., Fryer C.L., 2007, \apj 670, 747
\item[] Kouwenhoven M.B.N., Brown A.G.A., Zinnecker H., Kaper L., Portegies Zwart S.F., 2005, \aap 430, 137
\item[] Kouwenhoven M.B.N., Brown A.G.A., Portegies Zwart S.F., Kaper L., 2007, \aap 474, 77
\item[] Kraus S., Hofmann K.-H., Menten K.M., Schertl D., Weigelt G., Wyrowski F., Meilland A., Perraut K., Petrov R., Robbe-Dubois S., Schilke P., Testi L., 2010, Nat. 466, 339
\item[] Kroupa P., 2001, IAUS 200, 199
\item[] Krumholz M.R., Klein R.I., McKee C.F., Offner S.S.R., Cunningham A.J., 2009, Sci. 323, 754
\item[] Krumholz M.R., Thompson T.A., 2007, \apj 661, 1034
\item[] Kuiper R., Klahr H., Beuther H., Henning T., 2010, \apj 722, 1556
\item[] Larson R.B., 1978, \mnras 184, 69
\item[] Lef{\`e}vre, L., Marchenko, S.~V., Moffat, A.~F.~J., Acker, A., 2009, \aap, 507, 1141
\item[] Levato H., Malaroda S., Morrell N., Solivella G., 1987, \apjs 64, 487
\item[] Linder, N., Rauw, G., Sana, H., De Becker, M., Gosset, E. 2007, \aap, 474, 193
\item[] Lloyd Evans, T.\ 1979, \mnras, 186, 13
\item[] Luna, G.~J., Levato, H., Malaroda, S., Grosso, M., 2003, Information Bulletin on Variable Stars, 5375, 1
\item[] Ma\'{i}z-Apell\'{a}niz J., 2010, \aap 518, 1
\item[] Mason B.D., Gies D.R., Hartkopf W.I., Bagnuolo W.G., Ten Brummelaar T.A., McAlister H.A., 1998, \aj 115, 821
\item[] Mason B.D., Hartkopf W.I., Gies D.R., Henry T.J., Helsel J.W., 2009, \aj 137, 3358
\item[] Mason B.D., Ten Brummelaar T.A., Gies D.R., Hartkopf W.I., Thaller M.L., 1997, \aj, 114, 2112
\item[] McAlister H.A., Hartkopf W.I., Hutter D.J., Shara M.M., Franz O.G., 1987, \aj 92, 183
\item[] McAlister H.A., Mason B.D., Hartkopf W.I., Shara M.M., 1993, \aj 106, 1639
\item[] Miroshnichenko, A.~S.: 2010, {\it IAU Symp.} {\bf 272}, 304
\item[] Moffat A.F.J., 2008, IAUS 250, 119
\item[] Narayan R., Paczynski B., Piran T., 1992, \apj 395, L83
\item[] Niemela, V.~S., \& Gamen, R.~C., 2005, \mnras, 356, 974
\item[] Oudmaijer R.D., Parr A.M., 2010, \mnras 405, 2439
\item[] Paczynski B., 1991, Acta Astronomica 41, 257
\item[] Patel N.A., Curiel S., Sridharan T.K., Zhang Q., Hunter T.R., Ho, P.T.P., Torrelles J.M., Moran J.M., Gmez J.F., Anglada G., 2005, Nat. 437, 109
\item[] Penny, L.~R., Seyle, D., Gies, D.~R., et al., 2001, \apj, 548, 889
\item[] Preibisch T., Balega Y., Hofmann K.-H., Weigelt G., Zinnecker H., 1999, New Astronomy 4, 531
\item[] Rauw, G., Sana, H., Antokhin, I.~I., et al., 2001, \mnras, 326, 1149
\item[] Roberts L.C., Turner N.H., Ten Brummelaar T.A., 2007, \aj, 133, 545
\item[] Rubin, B.~C., Finger, M.~H., Harmon, B.~A., et al., 1996, \apj, 459, 259
\item[] Sana, H., Hensberge, H., Rauw, G., Gosset, E., 2003, \aap, 405, 1063
\item[] Sana, H., Gosset, E., Rauw, G.\ 2006, \mnras, 371, 67
\item[] Sana, H., Rauw, G., Gosset, E.\ 2007, \apj, 659, 1582
\item[] Sana H., Evans C.J., 2011, Proc. IAUS 272, 2011
\item[] Shepherd D.S., Claussen M.J., Kurtz S.E., 2001, Sci. 292, 1513
\item[] Schilbach E., R\"oser S., 2008, \aap 489, 105
\item[] Sota A., Ma\'{i}z-Apell\'{a}niz J., Walborn N.R., Shida R.Y., 2008, RMxAA 33, 56
\item[] Sota A., Ma\'{i}z-Apell\'{a}niz J., Walborn N.R., Alfaro E.J., Barb\'{a} R.H., Morrell, N.I., Gamen R.C., Arias J.I., 2011, \apjs 139, 24
\item[] Struve, O.\ 1944, \apj, 100, 189
\item[] Terrell, D., Munari, U., Zwitter, T., Nelson, R.~H., 2003, \aj, 126, 2988
\item[] van den Heuvel E.P.J., 2007, AIPC 924, 598
\item[] van Leeuwen, F., van Genderen, A.~M.\ 1997, \aap, 327, 1070
\item[] Walker, W.~S.~G., Marino, B.~F., 1972, Information Bulletin on Variable Stars, 681, 1
\item[] Wheelwright H.E., Oudmaijer R.D., Goodwin S.P., 2010, \mnras 401, 1199
 Shepherd D.S., Claussen M.J., Kurtz S.E., 2001, Sci. 292, 1513
\item[] Wolff S.C., 1978, \apj 222, 556
\item[] Yorke H., Sonnhalter C., 2002, \apj 569, 846 (2002)
\item[] Zinnecker H., Yorke H., 2007, AR\aap 45, 481

\end{itemize}

\end{document}